# An Accurate SER Estimation Method Based on Propagation Probability

Ghazanfar Asadi and Mehdi B. Tahoori
Northeastern University, Dept. of ECE, Boston MA 02115
Email :{gasadi, mtahoori}@ece.neu.edu

## Abstract

*In this paper, we present an accurate but very fast soft error rate (SER) estimation technique for digital circuits based on error propagation probability (EPP) computation. Experiments results and comparison of the results with the random simulation technique show that our proposed method is on average within 6% of the random simulation method and four to five orders of magnitude faster.*

## 1 Introduction

Soft errors are intermittent malfunctions of the hardware that are not reproducible [4]. These errors, which occur more often than permanent errors, arise from *Single Event Upsets* (SEU). These SEUs, in turn, arise from energetic particles, namely neutrons and alpha particles. *Soft Error Rate* (SER) for a device is defined as the error rate due to SEUs. So far, memory elements have been more susceptible to soft errors than the combinational logic. However, analytical models predict that the SER in the combinational logic will be comparable to that of memory elements by 2011 [6]. The first step in developing soft error reliable designs with minimum performance and area penalties is to accurately estimate system SER and the contribution of each component to the overall soft error vulnerability.

The error rate of a circuit node, $n_i$, is broken into three terms and computed as $R_{SEU}(n_i) \times P_{latched}(n_i) \times P_{sensitized}(n_i)$ [3]. $R_{SEU}(n_i)$ is the bit-flip rate at node $n_i$ which depends on the particle flux, the energy of the particle, type and size of the gate, and the device characteristics. $P_{latched}(n_i)$ is the probability that an erroneous value on node $n_i$ is captured in a flip-flop. $P_{sensitized}(n_i)$ is the probability that node $n_i$ is functionally synthesized by the input vectors to propagate the erroneous value from the error site to primary outputs (POs) or flip-flops (FFs). Estimating the last parameter is the most time-consuming part since in this step, several random vectors are applied to the circuit inputs to determine the *Error Propagation Probability* (EPP) from an error site to outputs. All previous SER estimation methods use the random vector simulation approach [2, 3, 4, 6]. The SER estimation time of a node in large circuits exponentially increases with the size of the circuit. Hence, SER estimation of larger circuits becomes intractable with these techniques.

In this paper, we present a new EPP computation technique based on circuit topological traversal and signal probabilities. This paper is organized as follows. Sec. 2 presents our EPP computation technique. Sec. 3 presents the experimental results. Finally Sec. 4 concludes the paper.

## 2 Gate-level SER Estimation

We consider all circuit nodes as possible error sites. In our approach, we first extract the structural paths from each error site to all reachable outputs. Then, we traverse these paths to compute the propagation probability of the erroneous value from the error site to reachable primary outputs or flip-flops. An *on-path* signal is a net on a path from the error site to a reachable output. An *on-path gate* is defined as the gate with at least one on-path input. Finally, an *off-path* signal is a net that is not on-path and is an input of an on-path gate. These three are shown in Fig. 1.

For EPP calculation, as we traverse the paths, we use signal probability for off-path signals and use our propagation probability rules for on-path signals. The signal probability (SP) of a line $l$ indicates the probability of $l$ having logic value "1" [5]. If there is only one path from the error site to an output, the error propagation probability from an on-path input of a gate to its output depends on the type of the gate and the signal probability of other off-path signals. In the general case in which reconvergent paths might exist, EPP from the error site to the output of the reconvergent gate depends on the polarities of the propagated error on the on-path signals, as well. To address this issue, we need error propagation rules for reconvergent gates. First, we define the following parameters:

- $P_a(U_i)$ and $P_{\bar{a}}(Ui)$ are defined as the probabilities of the output of node $U_i$ being $a$ and $\bar{a}$, respectively. ($a$ is an erroneous values and $\bar{a}$ is the inverted of $a$). In other words, $P_a(U_i)$ ($P_{\bar{a}}(Ui)$) is the probability that the erroneous value is propagated from the error site to $U_i$ with an even (odd) number of inversions.
- $P_1(U_i)$ and $P_0(U_i)$ are the probabilities of the output of node $U_i$ being 1 and 0, respectively. In these cases, the error is blocked and not propagated.

Note that for an on-path signal $U_i$, $P(U_i) = P_a(U_i) + P_{\bar{a}}(Ui) + P_1(U_i) + P_0(U_i) = 1$, while for an off-path signal $U_j$, $P(U_j) = P_1(U_i) + P_0(U_i) = 1$. Since we have considered the polarity of error propagation, this will take care of reconvergent fanouts. EPP computation rules for elementary gates are shown in Table 1.

Consider the example shown in Fig. 1. Assume that an SEU with sufficient energy hits gate $A$. After computing $P(E) = 1(\bar{a})$, $P(G) = 0.7(\bar{a}) + 0.3(0)$, and $P(D) = 0.2(a) + 0.8(0)$ [1] , we follow these steps for EPP calculation.

[1] This means: $P_a(D) = 0.2$, and $P_0(D) = 0.8$

| Gate | Rule |
|---|---|
| AND | $P_1(out) = \prod_{i=1}^{n} P_1(X_i)$<br>$P_a(out) = \prod_{i=1}^{n} [P_1(X_i) + P_a(X_i)] - P_1(out)$<br>$P_{\bar{a}}(out) = \prod_{i=1}^{n} [P_1(X_i) + P_{\bar{a}}(X_i)] - P_1(out)$<br>$P_0(out) = 1 - [P_1(out) + P_a(out) + P_{\bar{a}}(out)]$ |
| OR | $P_0(out) = \prod_{i=1}^{n} P_0(X_i)$<br>$P_a(out) = \prod_{i=1}^{n} [P_0(X_i) + P_a(X_i)] - P_0(out)$<br>$P_{\bar{a}}(out) = \prod_{i=1}^{n} [P_0(X_i) + P_{\bar{a}}(X_i)] - P_0(out)$<br>$P_1(out) = 1 - [P_0(out) + P_a(out) + P_{\bar{a}}(out)]$ |
| NOT | $P_1(out) = P_0(input)$, $P_a(out) = P_{\bar{a}}(input)$<br>$P_{\bar{a}}(out) = P_a(input)$ , $P_0(out) = P_1(input)$ |

**Table 1.** EPP calculation rules for elementary gates

$\mathbf{P_0(H)} = P_0(C) \times P_0(D) \times P_0(G) = 0.7 \times 0.8 \times 0.3 = 0.168$
$\mathbf{P_a(H)} = (0.7) \times (0.2 + 0.8) \times (0.3) - 0.168 = 0.042$
$\mathbf{P_{\bar{a}}(H)} = (0.7) \times (0.8) \times (0.7 + 0.3) - 0.168 = 0.392$
$\mathbf{P_1(H)} = 1 - (0.168 + 0.042 + 0.392) = 0.398$
$\rightarrow \mathbf{P(H)} = 0.042(a) + 0.392(\bar{a}) + 0.168(0) + 0.398(1)$

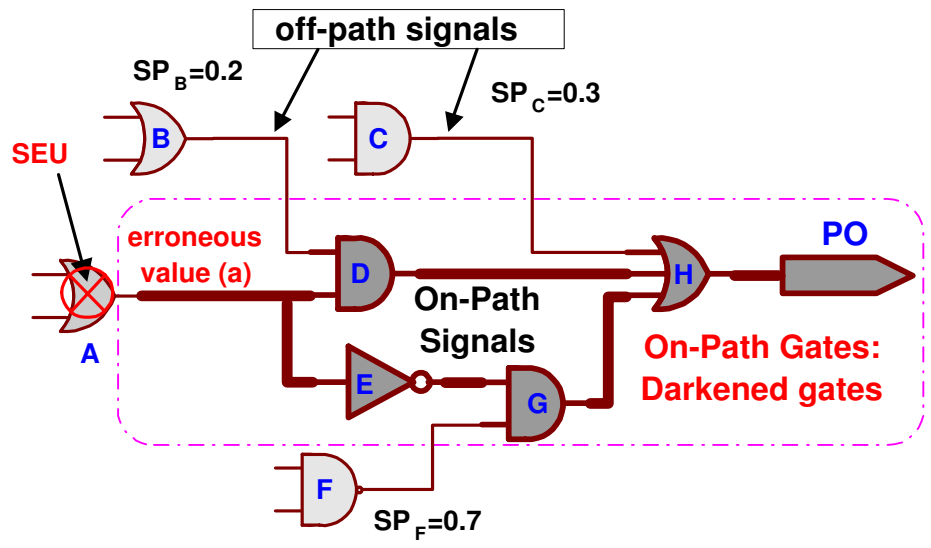

**Figure 1.** EPP calculation on reconvergent paths

For a general case, the following algorithm shows how we can extract and then traverse all paths from a given error site to all reachable outputs and how we apply the EPP rules as we traverse the paths.

For every node, $n_i$, do:

1. *Path Construction*: Extract all on-path signals (and gates) from $n_i$ to every reachable primary output $PO_j$ and/or flip-flop $FF_k$ using the forward *Depth-First Search* (DFS) algorithm [1].

2. *Ordering*: Levelize signals on these paths using the *topological sorting* algorithm [1].

3. *EPP Computation*: Select the on-path gates in a topological order, from the error site to reachable outputs, and apply propagation rules (Table 1) for EPP computation. Using a topological order enable us to compute EPP in just one pass (linear time complexity).

After completing the above steps, $P_a(PO_j)$ and $P_{\bar{a}}(PO_j)$ are computed for every $PO_j$ reachable from $n_i$. $P_{sensitized}(n_i)$ is calculated as follows:

$$P_{sensitized}(n_i) = \left(1 - \prod_{j=1}^{k} 1 - (P_a(PO_j) + P_{\bar{a}}(PO_j))\right)$$

, where $k$ is the number of outputs reachable from $n_i$.

## 3 Experimental Results

The proposed approach was implemented and applied to ISCAS'89 benchmark circuits using a DELL Precision

| Circuit | SysT | SimT | %Dif | SPT | ISP | ESP |
|---|---|---|---|---|---|---|
| s953 | 0.354 | 28.3 | 4.3 | 150 | 74.4 | 79950 |
| s1196 | 0.750 | 54.6 | 3.6 | 313 | 92.2 | 72800 |
| s1238 | 0.532 | 36.9 | 3.4 | 207 | 90.3 | 69510 |
| s1423 | 2.230 | 53.1 | 3.9 | 250 | 138.5 | 23810 |
| s1488 | 0.425 | 7.3 | 4.4 | 14 | 316.3 | 17220 |
| s1494 | 0.704 | 10.8 | 4.4 | 22 | 303.7 | 15480 |
| s9234 | 9.368 | 817.2 | 11.3 | 4659 | 970.8 | 87230 |
| s15850 | 34.18 | 972.1 | 12.6 | 5270 | 1695 | 28440 |
| s35932 | 7.020 | 1904 | 4.5 | 9648 | 3133 | 271240 |
| s38584 | 13.860 | 2317 | 7.1 | 12833 | 3405 | 167180 |
| s38417 | 14.180 | 2412 | 6.0 | 12951 | 3480 | 170126 |
| average | 3.243 | 325.0 | 5.4 | 110.7 | 549.1 | 93072 |

**Table 2.** Our approach vs. random simulation
SysT: Our approach run time(ms),SimT: Rand-Simul. run time(s)
Dif.: Difference of our approach vs. random simulation
SPT: Signal probability computation time (s)
ISP: Speedup including SP time, ESP: Speedup excluding SP time

450© system equipped with 2 GB memory. Table 2 shows the results for our systematic approach as well as the random simulation for selected circuits (due to space limitations). For larger circuits, a limited number of gates of the circuits are simulated due to exorbitant run time of the random-simulation method. The speedups of our approach are reported with and without including the SP computation time in the total execution time. When SP time is excluded, the speedups are 4-5 orders of magnitude. When included, our approach is still 2-3 orders of magnitude faster than the random simulation method. The accuracy of our approach versus random-simulation is 94%, in average.

## 4 Conclusions

In this paper, an accurate error propagation probability computation technique for soft error rate estimation has been developed. The proposed approach leverages the signal probability calculation, which is already used in other steps of the design flow. This technique can be used to identify the most vulnerable components to be protected by soft error hardening techniques.